\RequirePackage{fix-cm}
\documentclass[smallextended]{svjour3}       
\usepackage[italian,english]{babel}
\smartqed  
\usepackage{hyperref}
\usepackage{xspace}
\usepackage{amssymb}
\usepackage{amsmath}
\usepackage{lipsum}
\usepackage{dsfont}
\usepackage{multirow}
\usepackage{afterpage}
\usepackage{amsfonts}
\usepackage{bm}%
\usepackage{babel}
\usepackage{graphicx}
\usepackage{color}

 \usepackage{mathptmx}      
\usepackage{latexsym}
\begin{document}

\title{Characterizing quantum ensemble using geometric measure of quantum coherence 
}


\author{ R. Muthuganesan        \and
       V. K. Chandrasekar 
}


\institute{  Centre for Nonlinear Science \& Engineering, School of Electrical \& Electronics Engineering,
SASTRA Deemed University, Thanjavur, Tamil Nadu 613 401, India\at
              \email{rajendramuthu@gmail.com}          
            \and
         Centre for Nonlinear Science \& Engineering, School of Electrical \& Electronics Engineering,
SASTRA Deemed University, Thanjavur, Tamil Nadu 613 401, India\at
              \email{chandru25nld@gmail.com}
          }

\date{Received: date / Accepted: date}

\maketitle

\begin{abstract}

The characterization of the quantum ensemble is a fundamental issue in quantum  information theory and foundations. The ensemble is also useful for various quantum information processing. To characterize the quantum ensemble, in this article,  we generalize the coherence measure of a state to the quantum ensemble.  Exploiting  the fidelity and affinity between the ensemble, we propose a quantumness quantifier for the quantum ensemble. It is shown that the proposed  quantifier satisfies the necessary axioms of a bonafide measure of quantumness. Finally, we compute the quantumness of a few well-known ensembles. 

\end{abstract}

\section{Introduction}
\label{intro}
Quantum coherence, a direct consequence of the superposition principle \cite{Nielsen,Walls}, is fueled widespread attention at a fundamental level, as well as in terms of potential applications such as metrology \cite{Giovannetti2004,Demkowicz2014}, biology \cite{Sarovar2010,Lloyd2011,Huelga2013,Lambert2013}, thermodynamics,  nanoscale physics \cite{Aberg,Lostaglio}  and refrigeration \cite{Buffoni2019}. It is one of the fundamental characteristics of a quantum system that makes a clear distinction from the classical system can behave.  Recently, there has been a considerable effort devoted to quantifying quantum  coherence from a resource theoretic perspective. The resource theory of coherence consists of a set of axioms for function on a real valued function of quantum states which are introduced in Baumgratz et. al \cite{Baumgratz2014}. After this rigorous framework on coherence theory, various coherence measure has been introduced \cite{Girolami2014coh,Yu2017,Bu2017,Zhao2018,Zhu,Rana2016,Chen2018QINP,Wang2016,Jin,Xiong,Liu2017,Feng2017,Muthu2021}. 

In quantum theory, we have encountered the most situation with incomplete  prior information and this peculiar situation can be mathematically described by a quantum ensemble. It means that a quantum ensemble $\mathcal{E}=\{(p_i, \rho_i), i\in I \} $ which consists of a number of states $\rho_i$ ($i$ is an index) with
respective probabilities $p_i$ and $I$ is an index set. The states in the corresponding ensemble do not commute with each other and due to the noncommutativity, ensemble exhibits the quantum nature. Hence, the noncommutivity between the state is an important tool to characterize the quantum ensemble \cite{Li2017}. Further, the various quantum information processing such as quantum communication and cryptography demonstrated through the  quantumness (resources) of quantum ensemble instead of the quantum state. Hence, the quantum ensembles are both fundamental and practical objects in quantum information theory. Owing to the quantumness of ensemble as a physical resource, measures have been proposed from different perspectives such as relative entropy \cite{Luo2009}, commutators \cite{Li2017}, unitary similarity invariant norms \cite{Qi2018}, coherence \cite{Mao2019}, quantum correlations, accessible information, security of information transmission, and quantum cloning \cite{Luo2011,Ferro}.

The fidelity and affinity characterize the closeness or similarity between two states. These quantities are useful in comparing initial and final states. Hence, both measures useful in various quantum information processing such as quantum cloning \cite{Gisin1997}, teleportation \cite{Zhang2007},  metrology \cite{Yu}, quantum detection \cite{Gorin2006}, and estimation theory \cite{Gu2010}. Further, a monotonically decreasing function of fidelity and affinity is considered as a valid distance measure. Metrics on state space induced by fidelity and affinity are useful for quantification of geometric discord \cite{Muthuaffinity}, measurement-induced nonlocality \cite{MuthuMIN,MuthuMIN1}, and quantum coherence \cite{Xiong,Muthu2021}. In this article, we extend the definition of fidelity and affinity between state to the between ensemble. 

This paper is organized as follows. In Sec. \ref{Sec2}, we review the definition and its properties of fidelity and affinity.  In Sec. \ref{Sec3},  we generalize the definition of fidelity and affinity to the quantum ensemble. The quantumness measure for ensemble in terms of coherence is presented in \ref{sec4}.
Finally, in Sec. \ref{concl} we present the conclusions.
\section{Fidelity and Affinity of ensemble}\label{Sec2}
In order to characterize the quantum ensemble and quantify the quantumness of the ensemble, we first, recapitulate the definition and properties of quantum fidelity and affinity. Let $\mathcal{H}$ be the Hilbert space with dimension $d$ and $\mathcal{L}(\mathcal{H})$  consists of a set of density  matrices on $\mathcal{H}$. Assume $\rho, \sigma \in \mathcal{L}(\mathcal{H})$, then the fidelity between the states $\rho$ and $\sigma$ is defined as  \cite{Jozsa1994}
\begin{align}
F(\rho, \sigma)= \text{Tr}\sqrt{\rho \sqrt{\sigma}\rho}, 
\end{align}
which quantifies the closeness of two quantum states. Similar to fidelity, the affinity $A(\rho, \sigma)$ is also a measure of closeness between the states  $\rho$ and $\sigma$ and is defined as \cite{Kholevo1972,Luo2004}
\begin{align}
A(\rho, \sigma)= \text{Tr}\sqrt{\rho} \sqrt{\sigma}.
\end{align}
This definition is similar to the Bhattacharyya coefficient between two probability distributions (discrete or continuous) in classical probability theory \cite{Bhattacharyya}. 

 Here, $X(\sigma,\rho)$ refers to either  fidelity or affinity. We  now  list the fundamental  properties  of  the  $X(\sigma,\rho)$:
\begin{enumerate}
\item[($X$1)] $0\leq X(\rho,\sigma)\leq 1$ and $X(\rho,\sigma)=1$ if and only if $\rho=\sigma$. Moreover, $X(\rho,\sigma)=X(\sigma,\rho)$.

\item[($X$2)] $X(\rho,\sigma)$ is unitary invariant i.e., $X(\rho,\sigma)=X(U\rho U^{\dagger},U\sigma U^{\dagger})$ for any unitary operator $U$. 

\item[($X$3)] $X(\rho,\sigma)$ is multiplicative under tensor product:
\begin{align}
  X(\rho_1 \otimes \rho_2, \sigma_1\otimes \sigma_2)=X(\rho_1, \sigma_1) \cdot X(\rho_2, \sigma_2). \nonumber
\end{align}

\item[($X$4)] $X(\rho,\sigma)$ is monotonic under CPTP map $\Lambda$ i.e., $X(\Lambda(\rho),\Lambda(\sigma))\geq X(\rho,\sigma)$

\item[($X$5)]  For any orthogonal projectors $\Pi_i=|i\rangle \langle i|$, $X(\sum_i \Pi_i\rho\Pi_i,\sum_i\Pi_i\sigma\Pi_i)=\sum_iX(\Pi_i\rho\Pi_i,\Pi_i\sigma\Pi_i)$

\item[($X$6)] For a CPTP map,
\begin{align}
\sum_i X(K_i\rho K_i^{\dagger}, K_i \sigma K_i^{\dagger})=\sum_i X(p_i \rho_i, q_i \sigma_i)\geq  X(\rho, \sigma),
\end{align}
where $\rho_i=K_i\rho K_i^{\dagger}/p_i$ and $\sigma _i=K_i \sigma K_i^{\dagger}/q_i$ with $p_i=\text{Tr}(K_i\rho K_i^{\dagger})$ and $q_i=\text{Tr}(K_i \sigma K_i^{\dagger})$ are the respective probabilities after the super selection.
\end{enumerate}
The detailed proof of the above properties is given in Refs.\cite{Xiong,Muthu2021}. Next, we extend the notion of fidelity and affinity to the quantum ensemble. Consider, two quantum ensemble $\mathcal{E}=\{(p_i, \rho_i) : i\in I\} $  and $\mathcal{F}=\{(q_j, \sigma_j) : i\in J\} $  on the same Hilbert space. Then the fidelity or affinity between the ensemble is defined as 
\begin{align}
X(\mathcal{E},\mathcal{F})= ~^\text{inf}_{c} \sum_{i,j} c_{ij}X(\rho, \sigma),
\end{align}
where $X(\mathcal{E},\mathcal{F})$ refers to either fidelity or affinity between the ensembles and the infimum is taken over joint probability distribution $c=\{ c_{ij}\}$  with marginal distributions $\{ p_i\} $ and $\{q_j \} $. The fidelity or affinity between the ensemble has the following fundamental properties: 
\begin{enumerate}
 \item[i).] If $p_i=\delta_{ik}$ and $q_j=\delta_{jk}$ i.e.,two ensembles $\mathcal{E}$ and $\mathcal{F}$ degenerate to sets of single quantum state, $\mathcal{E}=\{1, \rho \} $,  $\mathcal{F}=\{1, \sigma \} $, then
 $X(\mathcal{E},\mathcal{F})$ reduces to conventional $X(\rho,\sigma)$ for any two singleton ensembles:
 \begin{align}
X(\mathcal{E},\mathcal{F})=X(\rho, \sigma).
\end{align}
 \item[ii).] $X(\mathcal{E},\mathcal{F})=1$  if and only if $\mathcal{E}$ and $\mathcal{F}$ are equivalent under trivial reductions. 
 \item[iii).] $X(\mathcal{E},\mathcal{F})$ is monotonic  under any completely positive and trace preserving (CPTP) maps.
\end{enumerate}

In general, both the fidelity and affinity are metric and one can define any monotonically decreasing function of $X(\rho,\sigma)$ as a metric in state space. One such  distance measure is 
\begin{align}
d_X(\rho,\sigma)=1-X(\rho,\sigma)
\end{align}
and we generalize the above distance measure to the quantum ensemble as 
\begin{align}
d_X(\mathcal{E},\mathcal{F})=1-X(\mathcal{E},\mathcal{F}).
\end{align}
In what follows, we extend the definition of coherence measure based on affinity and fidelity to the ensemble.


\section{Coherence of quantum ensemble}
\label{Sec3}
In this section, let us formally introduce the measures of coherence in the framework of resource theory. This resource theory is based on the set of incoherent operations as the free operations and the set of incoherent states as the set of free states. The set of incoherent states and the incoherent operations depend crucially on the choice of basis.  Firstly, we assume, $\{ |i\rangle, i=0,1, \cdots, d-1 \} $ be  a set of  orthonormal basis in the state space with finite dimension $d$. A  state is said to be an incoherent state if the corresponding density operator $\delta$ of the state is diagonal in this basis and denoted as $\delta=\sum_{i=0}^{d-1}\delta_i|i\rangle\langle i|$ with $\delta_i\in \mathbb{R}$ and $\sum_i \delta_i=1$. We label this set of incoherent quantum states as $\mathcal{I}$: 
\begin{align}
  \mathcal{I}=\left\{\delta =\sum_{i=0}^{d-1}\delta_i|i\rangle\langle i| \right\}.  \label{incohstate}
\end{align}
It is well-known that the incoherent state $\delta$ can have coherence in any other basis. All other states, which are not belonging to the above-mentioned set in the basis, are called coherent states.  In coherence theory, in view of the widely accepted notion of free states, there are different approaches to characterize the free operations. The free operations are named incoherent operations (IOs), referred to as the completely positive and trace-preserving (CPTP) maps that admit an incoherent Kraus representation \cite{Nielsen}. That is, there always exists a set of Kraus operators $\{ K_i\}$ such that 
\begin{align}
\Phi(\delta)\equiv \frac{K_i \delta K_i^{\dagger}}{\text{Tr}(K_i \delta K_i^{\dagger})} \in \mathcal{I}~; ~~~~~~ \sum_i K_i^{\dagger}K_i=\mathds{1}
\end{align}
for all $i$ and any  incoherent state $\delta$. These operations are called incoherent operations.

A functional $C(\rho)$ is a bonafide measure of quantum coherence of a state $\rho$, if it fulfills the following essential requirements \cite{Girolami2014coh,Baumgratz2014}. They are
\begin{enumerate}
\item[(C1)] \textit{Faithfulness}: $C(\rho) \geq 0$ is nonnegative and $C(\rho)=0$ if and only if $\rho$ is an incoherent state.

\item[(C2)] \textit{Monotonicity}: $C(\rho)$ does not increase under the action of an incoherent operation i.e., $C(\rho)\geq C(\Phi(\rho))$.

\item[(C3)] \textit{Strong Monotonicity}: $C(\rho)$ does not increase on average under selective incoherent operations, i.e., $C(\rho)\geq C(\sum_ip_i\rho_i))$, where $p_i=\text{Tr}(K_i \rho K_i^{\dagger})$ and $\rho_i= K_i \rho K_i^{\dagger}/p_i$ with $\sum_i K_i^{\dagger}K_i=\mathds{1}$.

\item[(C4)] \textit{Convexity}: Nonincreasing under mixing of quantum states i.e., $C(\sum_np_n\rho_n)\leq \sum_n p_nC(\rho_n)$ for any states $\rho_n$ with $\sum_np_n=1$.
\end{enumerate}

 More recently, the coherence measure based on fidelity and affinity has been introduced \cite{Xiong,Muthu2021} and the application of these coherence measures is also highlighted. The definition of the coherence measure is given as 
\begin{align}
C_X(\rho)=1- ~^\text{max}_{\delta \in \mathcal{I}} X(\rho, \delta),
\end{align}
where the optimization is taken over all possible incoherent states from the set as given in Eq. (\ref{incohstate}). It is shown that the fidelity measure could satisfy all the necessary axioms  (C1)-(C4) of a valid quantum resource. On the other hand, in general resource theory, affinity-based resource does not satisfy the convexity (C4) and not a valid resource. Recently, researchers have been shown that the affinity- based measure is a useful resource only in the framework of resource theory of coherence.  Without loss of generality, the above coherence measure rewritten as 
\begin{align}
C_X(\rho, \sigma)=1- ~^\text{max}_{\delta \in \mathcal{I}} X(\rho, \Pi(\rho)),
\end{align}
where $\Pi=\{\Pi_i =|i\rangle \langle i| \}$ is the von Neumann measurement corresponds to the given basis $|i\rangle $ and $\Pi(\rho)=\sum_i \Pi_i\rho \Pi_i$ is the post-measurement state.

In order to extend the above definition to the quantum ensemble,  first, we define the  incoherent ensembles $\Delta$, which can be identified as the ensemble freely access on the above mentioned basis, i.e., the ensemble that consists of only incoherent states (diagonal states). Similar to the coherence of a quantum state, the natural way to quantify the coherence of a given quantum ensemble $\mathcal{E}$ is the distance of this ensemble from the set of incoherent ensembles $\Delta$ in the reference basis. Exploiting affinity and fidelity between quantum ensemble we can propose a coherence quantifier for quantum ensemble  
\begin{align}
C_{X}(\mathcal{E}, \Delta)=~^\text{inf}_{\Delta}d_X(\mathcal{E},\Delta)=~^\text{~~~inf}_{\{ (q_j, \delta_j)\} }~^\text{inf}_{~c} \sum_{i,j} c_{ij} d_X(\rho_i, \sigma_j), \label{defen}
\end{align}
where $\Delta=\{(q_j ,\delta_j), j\in J\} $ is an incoherent ensemble, $\delta_j$ are the diagonal states in the considered basis $\{ |i\rangle, i=0,1, \cdots, d-1 \} $ and $c=\{c_{ij} \} $ is the joint probability distribution. The above quantity is quite hard to compute due to complex optimization. 

\textit{Theorem:1 The quantity $C_{X}(\mathcal{E})$ has the equivalent form},
\begin{align}
C_{X}(\mathcal{E})=\sum_i p_i C_X(\rho_i). \label{enmeasure}
\end{align}

Proof: Let $c=\{c_{ij} \} $ is the joint probability distribution with $c_{ij}=\sum_i p_i \delta_{ij}$. We have 
\begin{align}
C_{X}(\mathcal{E})=&\sum_i p_i C_X(\rho_i)=~^\text{~~inf}_{\delta_j \in \Delta} ~^\text{inf}_{~c}  \sum_{i,j} c_{ij}  d_X(\rho_i, \sigma_j)\\ \nonumber
 \leq& \sum_{i,j} p_i \delta_{ij} ~^\text{~~inf}_{\delta_j \in \Delta} d_X(\rho_i, \sigma_j)\\ \nonumber
=&\sum_{i} p_i ~^\text{~~inf}_{\delta_i \in \Delta} d_X(\rho_i, \sigma_i) \\ \nonumber
=&\sum_{i} p_i  C_X(\rho_i).
\end{align}
For the converse, we define the optimal incoherent ensemble $\tilde{\Delta}=\{(\tilde{q}_j, \tilde{\delta}_j), j\in J \}  $ and $\tilde{c}_{ij}$ be the optimal joint probability distribution with marginals distribution $\tilde{p}_{i}$ and $\tilde{q}_{j}$  which achieve the minimum (\ref{defen}). With these, one can define the nearest incoherent state of $\rho$ is $\tilde{\delta}_j=\Pi(\rho)=\sum_k|k\rangle \langle k|\rho|k\rangle \langle k|$. Then, we have 
\begin{align}
C_{X}(\mathcal{E})=\sum_i p_i C_X(\rho_i)  
= &~^\text{~~inf}_{\delta_j \in \Delta} ~^\text{inf}_{~c}   \sum_{i,j} c_{ij}  d_X(\rho_i, \sigma_j)  \\ \nonumber
=&\sum_{i,j} \tilde{c}_{ij}~d_X(\rho_i, \sigma_j), \\ \nonumber
\geq& \tilde{c}_{ij}~d_X(\rho_i, \Pi(\rho_i))   \\ \nonumber
=&\sum_i\left(\sum_j \tilde{c}_{ij} \right)C_X(\rho_i)=\sum_i p_i C_X(\rho_i),
\end{align}
which completes the proof of the theorem.  

The coherence measure for ensemble given in Eq. (\ref{enmeasure}) is  positive $C_{X}(\mathcal{E})\geq 0$   for any ensemble $\mathcal{E}$, monotonic under CPTP map and is convex in the sense that $C(\mathcal{E}) \leq  \sum_n t_n C(\mathcal{E}_n)$ with $t_n\geq 0$, $\sum_n t_n=1$ and $\mathcal{E}=\{(p_i, \rho_{in}), i\in I\}$ satisfies $\sum_n t_n\rho_{in}=\rho_i$. These properties are the direct consequence of the properties of $X(\rho, \sigma)$. Further, the measures fidelity and affinity are useful in quantification of coherence and distinguishability. Then the quantity $C_{X}(\mathcal{E})$ is interpreted as  (i) The minimal distinguishability between $\mathcal{E}$ and any incoherent ensembles, (ii) The average coherence of the constituent states in $\mathcal{E}$, (iii) The coherence of the bipartite state with a label space i.e., $\sum_i p_i|i\rangle \langle i|\otimes \rho_i$. 

\section{Quantumness of ensembles}\label{sec4}
In this section, we introduce a new quantifier of quantumness in ensembles from the perspectives of the quantum coherence measure identified  above. The quantumness of the ensemble is regarded as a useful kind of quantum resource for the various applications, in particular, quantum cryptography.   Before proceeding further, we introduce the terminologies (free ensembles and free operations) used  in the resource theory for the ensemble. The free ensembles are the ensembles that possess no quantumness, which can be naturally identified as classical ensembles (diagonal) in the sense that all the constitute states commute with each other. Further, it is important to highlight that free ensembles are the basis-independent version of incoherent ensembles, since incoherent ensembles are classical and all classical ensembles are incoherent in a specific basis. The free operations for quantumness can be characterized as the commutativity preserving operations (CPOs), which map all classical ensembles to classical ensembles. For the given ensemble $ \mathcal{E}=\{(p_i, \rho_{i}), i\in I\}$, the measure of quantumness in the ensemble is defined as 
\begin{align}
Q_{X}(\mathcal{E})=~^\text{min}_{~U} ~^\text{inf}_{\Delta}~ C_X(\mathcal{E}, U\Delta U^{\dagger}), \label{Qmeasure}
\end{align}
where $U$ is the arbitrary unitary operator and $\Delta$ is the incoherent ensemble in a preferred  basis. In general,  a bonafide quantifier of quantumness monotone should at least satisfy the following three conditions: (i) vanishes only for the classical ensemble, (ii) nonincreasing under CPOs, (iii) invariant under unitary operations, since noncommutativity can not be altered by unitary operations. 

To show the quantumness for the ensemble as a valid measure, we list the fundamental properties of $Q_{X}(\mathcal{E})$ given in Eq. (\ref{Qmeasure})
\begin{enumerate}
  \item[(i)] Positivity: $Q_{X}(\mathcal{E})\geq $ for any quantum ensembles $\mathcal{E}$, with equality if and only if $\mathcal{E}$ is a classical ensemble (diagonal state) in the sense that all its constituent states commute with each other. In particular, the ensemble is singleton, its quantumness is zero.
  
  \item[(ii)] Unitary invariance: $Q_{X}(\mathcal{E})$ is invariant under unitary transformation in the sense that $Q_{X}(\mathcal{E})=Q_{X}(U\mathcal{E}U^{\dagger})$ for any unitary operator $U$ with $U\mathcal{E}U^{\dagger}=\{ (p_i, U\mathcal{E}U^{\dagger}), i \in I\}$.

   \item[(iii)] Monotonicity under any commutativity preserving operations  $Q_{X}(\mathcal{E})$ is nonincreasing  under any commutativity preserving operation $M$ i.e., $Q_{X}(\mathcal{E})\geq Q_{X}(M(\mathcal{E}))$, where $M(\mathcal{E})=\{ (p_i, M(\mathcal{E}), i \in I\}$.
   
   \item[(iv)] Concavity under probabilistic union: $Q_{X}(\mathcal{E})$ is concave in the sense that $Q\left(\bigcup_{\mu} \lambda_{\mu}\mathcal{E}_{\mu}\right)\geq \sum_{\mu}\lambda_{\mu}Q(\mathcal{E}_{\mu}) $. Here, $\lambda_{\mu}\geq 0$, $\sum_{\mu \in K} \lambda_{\mu}=1$ and for each $\mu \in K$, $\mathcal{E}_{\mu}=\{(p_{\mu i}, \rho_{\mu i}), i\in I_{\mu} \} $ is a quantum ensemble and $\bigcup_{\mu} \lambda_{\mu}\mathcal{E}_{\mu}=\{(\lambda_{\mu}p_{\mu i}, \rho_{\mu i}), \mu \in K i\in I_{\mu}\}$ is a probabilistic union of $\mathcal{E}_{\mu}$.

  \item[(v)] Concavity of probabilistic distributions: $Q_{X}(\mathcal{E})$ is concave in the sense that $Q_{X}(\mathcal{E})\geq \sum_n t_nQ_{X}(\mathcal{E}_n)$ with $t_n\geq 0$,  $t_n=1$ and $ \mathcal{E}=\{(p^{(n)}_i, \rho_{i}), i\in I\}$ satisfies $\sum_nt_np^{(n)}_i=p_i$. Implying that the ensembles  $ \mathcal{E}_n$ are probability distribution decomposition of ensemble $ \mathcal{E}$ with fixed constituent states.
   
  \item[(vi)] Increasing under fine-graining: $Q_{X}(\mathcal{E})$ is increasing under fine-graining in the sense that $Q_{X}(\mathcal{E})\leq Q_{X}(\mathcal{E}_F)$. Here, $ \mathcal{E}=\{(p_i, \rho_{i}), i\in I\}$ with each constituent state $\rho_i$ being decomposed into a new quantum ensemble $ \mathcal{E}_F=\{(\lambda_{i\mu}, \rho_{i\mu}), i\in I, \mu\in J_i\}$ is the fine-grained ensemble such that $\rho_i=\sum_{\mu\in J_i}\lambda_{i\mu}, \rho_{i\mu}$.
   
   \item[(vii)]  Decreasing under coarse-graining: $Q_{X}(\mathcal{E})\geq Q_{X}(\mathcal{E}_C)$. Here $ \mathcal{E}=\{(p_i, \rho_{i}), i\in I\}$ and $\{c_s, s\in S \} $ is a partition of the index set $I$ i.e., $\bigcup_s c_s=I, c_s\cap c_t=\emptyset $ for $s\neq t$, $\mathcal{E}_C=\{p_{c_s}, \rho_{c_s}, s\in S\} $ is the coarse-grained ensemble with $p_{c_s}=\sum_{i\in c_s}p_i$ and $\rho_{c_s}=(p_{c_s})^{-1}\sum_{i\in c_s}p_i\rho_i$.
\end{enumerate}

Here we provide proof of all the properties of quantumness measure of the ensemble. The properties (i) and (ii) directly follow from positivity and unitary invariance of $C_X(\rho)$. To prove (iii), assume $\tilde{\Delta}=\{(\tilde{q}_j, U\tilde{\delta}_jU^{\dagger}), j\in J \}  $ the optimal incoherent ensemble and $\tilde{c}_{ij}$ be the optimal joint probability distribution with marginals distribution $\tilde{p}_{i}$ and $\tilde{q}_{j}$, which makes Eq. (\ref{Qmeasure}) is minimum. Then we have, 
\begin{align}
Q_{X}(\mathcal{E})=~^\text{min}_{~U} ~^\text{inf}_{\Delta}~ C_X(\mathcal{E}, U\Delta U^{\dagger})=\sum_{i,j}\tilde{c}_{ij}C_X(\rho_i, U\tilde{\delta}_i U^{\dagger}).
\end{align}
Since $X(\cdot)$ is monotonic under CPTP map and M preserves commutativity, a classical state is mapped to a classical one. Then the $Q_{X}(\mathcal{E})$ is 
\begin{align}
Q_{X}(\mathcal{E})\geq &\sum_{i,j}\tilde{c}_{ij}C_X(M(\rho_i), M(U\tilde{\delta}_i U^{\dagger})), \\ \nonumber
\geq& ~^\text{min}_{~U} ~^\text{inf}_{\Delta} C_X(M(\rho_i), U\Delta U^{\dagger}) \\ \nonumber
=&C_X(M(\mathcal{E})).
\end{align}
The above inequality can be derived from the minimum. Hence, the above three properties validate the $Q_{X}(\mathcal{E})$ is a bonafide quantifier of quantumness of ensemble $\mathcal{E}$. 

To show the property (iv), assume that $U_{\mu}$ and $U_\mathcal{E}$ are the particular unitary operators that achieve the minimum in Eq. (\ref{Qmeasure}) for each corresponding quantum ensemble $\mathcal{E}_{\mu}$ and $\mathcal{E}$ respectively. Then 
\begin{align}
\sum_{\mu}\lambda_{\mu}Q(\mathcal{E}_{\mu})=&\sum_{\mu}\lambda_{\mu} \sum_i p_{i\mu}C_X(U^{\dagger}_{\mu}\rho_{i\mu}U_{\mu}) \\ \nonumber
\leq &\sum_{\mu}\lambda_{\mu} \sum_i p_{i\mu}C_X(U^{\dagger}_{\mathcal{E}}\rho_{i\mu}U_{\mathcal{E}}) \\ \nonumber
=&\sum_{i, \mu} \lambda_{\mu} C_X(U^{\dagger}_{\mathcal{E}}\rho_{i\mu}U_{\mathcal{E}}) \\ \nonumber
=&Q(\mathcal{E}).
\end{align}

To establish property (v), again we assume that the $\tilde{c}_{ij}$ be the optimal joint probability distribution  and $\delta'=\{\tilde{U}\tilde{\delta}\tilde{U^{\dagger}} \}$ be the set of commuting states which achieve the minimum in Eq. (\ref{Qmeasure}). The marginal distributions $\{p^{(n)}_i \} $ satisfying that $\sum_ns_n\tilde{c}_{ij}^{(n)}=\tilde{c}_{ij}, ~\forall ~i, j$. Then 
\begin{align}
Q(\mathcal{E})=& ~^\text{min}_{~U} ~^\text{inf}_{\Delta}~ C_X(\mathcal{E}, U\Delta U^{\dagger}) \nonumber \\ 
= &\sum_{i,j} \tilde{c}_{ij} C_X(\rho_i, \tilde{U}\tilde{\delta}_j\tilde{U}^{\dagger}) \nonumber \\
=&\sum_{i,j} \sum_n s_n\tilde{c}_{ij}^{(n)} C_X(\rho_i, \tilde{U}\tilde{\delta}_j\tilde{U}^{\dagger}).  \nonumber
\end{align}
Interchanging the summation, we have
\begin{align}
Q(\mathcal{E})=&\sum_n s_n \sum_{i,j} \tilde{c}_{ij}^{(n)} C_X(\rho_i, \tilde{U}\tilde{\delta}_j\tilde{U}^{\dagger}) \nonumber \\ 
\geq &\sum_ns_n ~^\text{min}_{~U} ~^\text{inf}_{\Delta}~ C_X(\mathcal{E}, U\Delta U^{\dagger})=\sum_n s_n Q(\mathcal{E}_n).
\end{align}

The property (vi) can be derived using the convexity of relative entropy coherence of quantum states. Let $U_F$ be the particular unitary operator that achieves the minimum in
Eq. (\ref{Qmeasure}) for $\mathcal{E}$, we have
\begin{align}
Q(\mathcal{E})=& ~^\text{min}_{~U} \sum_i p_i C_X(U^{\dagger}\rho_iU) \nonumber \\ 
\leq &\sum_i p_i C_X(U^{\dagger}_{\mathcal{E}_F}\rho_iU_{\mathcal{E}_F}) \nonumber \\
=&\sum_i p_iC_X(U^{\dagger}_{\mathcal{E}_F} (\sum_{\mu\in J_i}\lambda_{i\mu} \rho_{i\mu})  U_{\mathcal{E}_F}) \nonumber \\
\leq& \sum_i p_i\sum_{\mu\in J_i}\lambda_{i\mu}C_X(U^{\dagger}_{\mathcal{E}_F}\rho_{i\mu}U_{\mathcal{E}_F}) \nonumber \\
=&\sum_{i,\mu}p_i\lambda_{i\mu}C_X(U^{\dagger}_{\mathcal{E}_F}\rho_{i\mu}U_{\mathcal{E}_F})=Q(\mathcal{E}_F).
\end{align}

Similarly, one can show the property (vii) of  $Q(\mathcal{E})$.
\vspace{0.5cm}

Next, we compute the quantumness of the following ensemble, which are used in different information processing such as quantum key distribution and cryptography. 
\begin{itemize}
  \item[(i)] $\mathcal{E}_{B92}=\{(1/2,|0\rangle), (1/2,|\psi_{\pi/4}\rangle)  \} $ is the ensemble used in the B92 quantum key distribution protocol with $|\psi\rangle=\text{sin}\theta|0\rangle+\text{cos}\theta|1\rangle$ \cite{Bennett1992}.
  The quantumness are $Q_{\mathcal{A}}(\mathcal{E})=0.293$ and $Q_{\mathcal{F}}(\mathcal{E})=0.5$.
  \item[(ii)] The ensemble utilized in the BB84 quantum key  distribution protocol is $\mathcal{E}_{BB84}=\{(1/4,|0\rangle), (1/4,|1\rangle),\\(1/4,|\psi_{\pi/4}\rangle),(1/4,|\psi_{3\pi/4}\rangle ) \} $ \cite{Bennett1984}.
   The quantumness of the ensemble are computed as $Q_{\mathcal{A}}(\mathcal{E})=0.293$ and $Q_{\mathcal{F}}(\mathcal{E})=0.5$
  \item[(iii)] The trine ensemble $\mathcal{E}_{trine}=\{ (1/3,|0\rangle), (1/3,(|0\rangle+\sqrt{3}|1\rangle)/2),(1/3,(|0\rangle-\sqrt{3}|1\rangle)/2)\} $ is useful in PBC00 protocol in quantum key distribution\cite{Phonex,Boileau}.
  The quantumness are $Q_{\mathcal{A}}(\mathcal{E})=0.293$ and $Q_{\mathcal{F}}(\mathcal{E})=0.5$.
 \item[(iv)] The mutually unbiased bases ensemble $\{(1/6, |\phi_i\rangle ),(1/6, |\phi_j\rangle ) \} $ with $i,j=x,y,z$ $|\phi_{i,j}\rangle$ are eigenvectors of Pauli spin matrices. The above ensemble useful in six-state protocol in quantum cryptography \cite{Bru}.
 The quantumness of the  ensemble are computed as $Q_{\mathcal{A}}(\mathcal{E})=0.293$ and $Q_{\mathcal{F}}(\mathcal{E})=0.5$
\end{itemize}

\vspace{1cm}

\section{Conclusion}
\label{concl}
To summarize, we have generalized the definition of geometric coherence measures of a quantum state based on fidelity and affinity to the quantum ensemble. Extending the coherence resource theory of the state, we set up a framework for the quantification of resource content in the quantum ensemble. The coherence measure of the ensemble is interpreted as minimal distinguishability and average coherence of ensemble.  Further, we have defined the minimal coherence of the ensemble in all bases as the quantumness measure. It is shown that the quantumness measure of ensemble possesses some important  properties such as positivity, unitary invariance and non-increasing under CPOs and has been proven as a bonafide measure of quantumness in quantum ensembles.   




%
%

\begin{acknowledgements}
This work financially supported by the Council of Scientific and Industrial Research (CSIR), Government of India, under Grant No. 03(1444)/18/EMR-II.
\end{acknowledgements}



\end{document}